\documentclass[prb,aps,twocolumn,amsmath,amssymb]{revtex4-2}
\usepackage{graphicx,bm}
\usepackage{dcolumn}

\begin{document}

\title{Quantum Field Theory of Correlated Bose-Einstein condensates: \\
I. Basic Formalism}

\author{Takafumi K{\sc ita}}
\affiliation{Department of Physics, Hokkaido University, Sapporo 060-0810, Japan}

\begin{abstract}
Quantum field theory of equilibrium and nonequilibrium Bose-Einstein condensates is formulated 
so as to satisfy three basic requirements: the Hugenholtz-Pines relation;
conservation laws; identities among vertices originating from Goldstone's theorem I.
The key inputs are irreducible four-point vertices, in terms of which we derive a closed system of equations
for Green's functions, three- and four-point vertices, and two-particle Green's functions.
It enables us to study correlated Bose-Einstein condensates with a gapless branch of single-particle excitations
without encountering any infrared divergence.
The single- and two-particle Green's functions are found to share poles,
i.e., the structure of the two-particle Green's functions predicted by Gavoret and Nozi\`eres for a homogeneous condensate at $T\!=\!0$ is also shown to 
persist at finite temperatures, in the presence of inhomogeneity, and also in nonequilibrium situations. 
\end{abstract}

\maketitle

\section{Introduction}
The aim of this paper is to derive a closed system of self-consistency equations for 
the single- and two-particle Green's functions of correlated Bose-Einstein condensates,
which is formally exact and can also be used in practical calculations to describe both equilibrium and nonequilibrium condensates.
This will be performed in such a way that it automatically meets the three exact requirements: 
(i) the Hugenholtz-Pines relation predicting a branch of gapless single-particle excitations \cite{HP59};
(ii) conservation laws \cite{KB62,Baym61,Baym62};
(iii) identities among vertices \cite{CCPS97,Kita19-1} originating from Goldstone's theorem I,
i.e., the first proof \cite{GSW62,Weinberg96}.
 
We have already made similar attempts in terms of (i) and (ii) \cite{Kita09,Kita10,Kita14}.
However, the resulting self-consistent perturbation expansion has encountered an infrared divergence,
similarly as in the case of simple perturbation expansion \cite{GN64} starting from either the ideal gas or the Bogoliubov theory \cite{Bogoliubov47}.
which has prevented us from performing practical calculations on correlated Bose-Einstein condensates.
A key additional observation here, which originates from our previous renormalization-group study \cite{Kita19-1,Kita19-2},
is that the infrared divergence can only be removed
by extending the self-consistency procedure beyond the self-energies up to the four-point vertices 
so as to satisfy hierarchical identities among two-, three-, and four-point vertices \cite{CCPS97,Kita19-1} as dictated by Goldstone's theorem I \cite{GSW62,Weinberg96}.

The background of the present study is briefly sketched as follows.
Bogoliubov \cite{Bogoliubov47} pioneered a microscopic description of interacting Bose-Einstein condensates 
to predict that the quadratic energy-momentum relation of free particles should be changed upon switching on the interaction 
into a linear sound-wave-like dispersion, whose speed is proportional to the square root of the bare interaction $U_0$.
Beliaev \cite{Beliaev58} formulated a field-theoretic perturbation expansion in terms of Green's functions.
Hugenholtz and Pines \cite{HP59} proved that single-particle excitations should have a gapless branch.
Gavoret and Nozi\`eres \cite{GN64} performed a structural analysis of the perturbation expansion
for the single- and two-particle Green's functions
to show that they have a common branch of poles. 
Nepomnyashchi\u{i} and Nepomnyashchi\u{i} \cite{Nepomnyashchii75,Nepomnyashchii78} 
used the identity between the two- and three-point vertices derived by Gavoret and Nozi\`eres \cite{GN64} to conclude
that the anomalous self-energy
should vanish in the low energy-momentum limit, contrary to the Bogoliubov theory where it is finite and proportional to the bare interaction $U_0$.
These basic studies consider only homogeneous Bose-Einstein condensates in equilibrium at $T\!=\!0$.
The field-theoretic approach has also encountered difficulties in practical applications 
such as the infrared divergence mentioned above 
or the conserving-gapless dilemma \cite{HM65,Griffin96}.

The present formulation covers both equilibrium condensates at finite temperatures 
and nonequilibrium ones. 
It will proceed by combining Schwinger's functional derivative method based on the generating functional \cite{Schwinger51,KB62,NO88,Swanson92},
the Legendre transformation to the effective action \cite{Weinberg96,dDM64,JL64,NO88,Swanson92},
the Luttinger-Ward functional \cite{LW60}, and conserving gapless condition \cite{Kita09, Kita14}.
A similar approach was adopted previously to analyze properties of two-particle Green's functions \cite{Kita10},
which however reached an erroneous conclusion that the single- and two-particle Green's functions do not have common poles.
It will be reexamined here by (i) incorporating the identities among the vertices 
and (ii) correcting the form of the perturbation.
The resulting revised conclusion is that the single- and two-particle Green's functions do share poles
not only at $T\!=\!0$ of a homogeneous condensate, as predicted by  Gavoret and Nozi\`eres \cite{GN64} and also restated recently by Watabe \cite{Watabe20}, but also at finite temperatures, in the presence of inhomogeneity, and also in nonequilibrium situations. 
As a bonus, we will be able to clarify the connections among the vertices which were not given in the Gavoret-Nozi\`eres study \cite{GN64}.

This paper is organized as follows.
Section \ref{sec2} studies properties of the condensate wave function $\Psi$ and Green's functions $G$ in equilibrium in terms of the effective action.
Section \ref{sec3} derives expressions of the three-point and four-point (i.e., two-particle) Green's functions based on the functional derivative method.
Section \ref{sec4} obtains self-energies in terms of $\Psi$, $G$, and vertices.
Section \ref{sec5} summarizes the key equations derived and also supplement them with
equations for the irreducible four-point vertices to construct a closed system of equations.
Section \ref{sec6} performs a nonequilibrium extension.
Section \ref{sec7} presents concluding remarks.

\section{Effective Action and Green's Functions\label{sec2}}

\subsection{System and Partition Function}

We consider a system of identical bosons with mass $m$ and spin $0$ 
described by the dimensionless action \cite{Weinberg96,Swanson92,NO88}
\begin{equation}
S=S_0+S_{\rm int},
\label{S}
\end{equation}
with 
\begin{subequations}
\label{S_0-S_int}
\begin{align}
S_0=&\,\int   d x\,\psi^*(x)\!\left(\frac{\partial}{\partial\tau}+\frac{\hat{\bf p}^2}{2m} -\mu\right)\!\psi(x) ,
\label{S_0}
\\
S_{\rm int}=&\,\frac{1}{2}\int  d x\int  d x'\,
U(x,x')\psi^*(x)\psi^*(x')\psi(x')\psi(x).
\label{S_int}
\end{align}
\end{subequations}
Here $\psi$ is the complex bosonic field and $\psi^*$ its conjugate,
$x\equiv ({\bf r},\tau)$ specifies a space-``time'' point with $0\!\leq\! \tau \!\leq\! \beta\equiv (k_{\rm B}T)^{-1}$
($k_{\rm B}$:  Boltzmann constant,  $T$: temperature), $\hat{\bf p}\equiv -i\hbar{\bm\nabla}$ is the momentum operator, $\mu$ is the chemical potential,
and 
\begin{align}
U(x,x')\equiv\delta(\tau-\tau')U(|{\bf r}-{\bf r}'|)
\end{align}
is the interaction potential.
We regard $\psi(x)$ and $\psi^*(x)$ as elements of a column vector, 
\begin{align}
\begin{bmatrix}
\psi(x) \\ \psi^*(x)
\end{bmatrix}
\equiv \begin{bmatrix}
\psi_1(x) \\ \psi_2(x)
\end{bmatrix}\equiv \vec{\psi}(x),
\label{vecpsi}
\end{align}
and will often express $\psi_j(x)\equiv \psi(\xi)$ with $\xi\equiv(j,x)$ and $j=1,2$.

Next, we introduce the grand partition function $Z_{JI}\!\equiv\!Z[J,I]$ with extra source functions \cite{Schwinger51}
$J(\xi)$ and $I(\xi,\xi')$ by
\begin{align}
Z_{JI}
\equiv&\,  \int D[\psi] \, \exp\left[-S-\int d\xi\,\psi (\xi) J(\xi)\right.
\notag \\
&\,
\left.-\int d\xi\int d\xi'\psi(\xi)\psi(\xi') I(\xi',\xi)\right],
\label{Z}
\end{align}
with $\displaystyle\int d\xi\equiv\sum_j\int dx$.
It satisfies
\begin{subequations}
\label{dlnZ}
\begin{align}
-\frac{\delta \ln Z_{JI}}{\delta J(\xi)}=&\,\langle \psi(\xi)\rangle_{JI}\equiv \Psi_{JI}(\xi),
\label{dlnZ1}
\\
-\frac{\delta \ln Z_{JI}}{\delta J(\xi)\delta J(\xi')}=&\,-\langle T_\tau\psi(\xi)\psi(\xi')\rangle_{JI}
+\Psi_{JI}(\xi)\Psi_{JI}(\xi')
\notag \\
\equiv &\, G_{JI}(\xi,\xi'),
\label{dlnZ11}
\\
-\frac{\delta \ln Z_{JI}}{\delta I(\xi',\xi)}=&\,\langle T_\tau \psi(\xi)\psi(\xi')\rangle_{JI}
\notag \\
=&\,  
-G_{JI}(\xi,\xi')+\Psi_{JI}(\xi)\Psi_{JI}(\xi') 
\notag \\
\equiv &\, -{\cal G}_{JI}(\xi,\xi'),
\label{dlnZ2}
\end{align}
\end{subequations}
where $T_\tau$ is the ``time''-ordering operator \cite{AGD63} and subscript $_{JI}$ emphasizes that  $J$ and  $I$ are  finite.

Introduction of the two-point external source function $I(\xi,\xi')$, besides $J(\xi)$ in the standard formalism \cite{Weinberg96,Swanson92,NO88}, is one of the key ingredients here.
Indeed, it enables us to express the effective action in terms of the renormalized Green's function
$G(\xi,\xi')$ instead of the bare propagator $G_0(\xi,\xi')$, as seen below.

\subsection{Effective Action}

Let us perform a Legendre transformation from $-\ln Z_{JI}$ into the effective action \cite{Weinberg96,dDM64,JL64,Swanson92,NO88}
\begin{align}
\Gamma_{JI}\equiv&\, -\ln Z_{JI} -\int  d\xi\,\Psi_{JI}(\xi) J(\xi)
\notag \\
&\, +\int  d \xi\int d\xi'\,{\cal G}_{JI}(\xi,\xi')I(\xi',\xi),
\label{Gamma}
\end{align}
which is a functional of $(\Psi_{JI},G_{JI})$. 
Its first derivatives with respect to $\Psi_{JI}$ and $G_{JI}$ can be calculated by considering 
their explicit dependences only;
the implicit dependences through $(J,I)$ cancel out because of Eq.\ (\ref{dlnZ}). 
Thus, we obtain
\begin{subequations}
\label{Gamma-deriv}
\begin{align}
\frac{\delta \Gamma_{JI}}{\delta \Psi_{JI}(\xi)}
=&\,-J(\xi)-\int  d\xi' \left[I(\xi,\xi')+I(\xi',\xi)\right]
 \Psi_{JI}(\xi'),
\label{Gamma-deriv1}
\end{align}
\begin{align}
\frac{\delta \Gamma_{JI}}{\delta G_{JI}(\xi',\xi)}=&\, \frac{I(\xi',\xi)+I(\xi,\xi')}{2},
\label{Gamma-deriv2}
\end{align}
\end{subequations}
where we have incorporated the symmetry $G_{JI}(\xi',\xi)\!=\!G_{JI}(\xi,\xi')$ in the second differentiation.

Next, we introduce the functionals
\begin{subequations}
\label{Gamma_J-Gamma-def}
\begin{align}
\Gamma_{J}\equiv &\,\Gamma_{JI}\bigr|_{I=0},
\\
\Gamma \equiv &\, \Gamma_{J}\bigr|_{J=0},
\end{align}
\end{subequations}
and correspondingly, $(\Psi_{J},G_{J})$ and $(\Psi,G)$.
Functionals $\Gamma_{J}$ and $\Gamma$ satisfy Eq.\ (\ref{Gamma-deriv}) with $I\!=\!0$ and $I\!=\!J\!=\!0$, respectively.
Thus, $\Gamma=\Gamma[\Psi,G] $ obeys
\begin{subequations}
\label{dGamma}
\begin{align}
\frac{\delta \Gamma}{\delta \Psi(\xi)}= &\,0,
\label{dGamma1}
\\
\frac{\delta \Gamma}{\delta G(\xi',\xi)}=&\,0,
\label{dGamma2}
\end{align}
\end{subequations}
which determine $(\Psi,G)$ in equilibrium.
Indeed, $\Gamma$ is connected with the grand potential $\Omega$ in equilibrium
by $\Gamma= \beta \Omega$.

It should be noted that $G_{J}$ in $\Gamma_{J}$ ($G$ in $\Gamma$) is a functional of $\Psi_{J}$ ($\Psi$), i.e.,
$G_{J}=G_J[\Psi_J]$ ($G=G[\Psi]$),
unlike the case of $\Gamma_{JI}$ where
$G_{JI}$ is independent of $\Psi_{JI}$. 
On the other hand, it also follows from Eq.\ (\ref{dGamma}) that $G$ in $\Gamma$ can be regarded as independent of $\Psi$ up to the linear order.
Put it another way, the total derivative $\delta$ in Eq.\ (\ref{dGamma}) can be replaced by the partial derivative,
which we will express by $\partial$.

Let us expand $\Gamma_{J}=\Gamma_{J}[\Psi_J]$ formally with respect to
\begin{align}
\delta \Psi(\xi)\equiv \Psi_{J}(\xi)-\Psi(\xi) 
\label{dPsi}
\end{align}
in the Taylor series
\begin{align}
\Gamma_{J}=&\,\Gamma+\sum_{n=1}^\infty \int d\xi_1\cdots\int d\xi_n\frac{\Gamma^{(n)}(\xi_1,\cdots,\xi_n)}{n!}
\notag \\
&\,\times
\delta\Psi(\xi_1)\cdots
\delta\Psi(\xi_n) .
\label{Gamma_J-exp}
\end{align}
It follows from Eq.\ (\ref{dPsi}) that $\Gamma^{(n)}$ can be written in terms of 
$\Gamma$ in equilibrium by
\begin{align}
\Gamma^{(n)}(\xi_1,\cdots,\xi_n)=\frac{\delta^n \Gamma}{\delta\Psi(\xi_1)\cdots\delta\Psi(\xi_n)} .
\label{Gamma^(n)-def}
\end{align}
Thus, Eq.\ (\ref{dGamma1}) is expressible alternatively as 
\begin{subequations}
\label{Gamma^(1,2)}
\begin{align}
\Gamma^{(1)}(\xi)=0 .
\label{Gamma^(1)=0}
\end{align}
In addition, $\Gamma^{(2)}$ satisfies \cite{Weinberg96,Swanson92,NO88}
\begin{align}
\Gamma^{(2)}(\xi,\xi')=&\,-G^{-1}(\xi,\xi') .
\label{Gamma^(2)}
\end{align}
\end{subequations}
This can be shown by (i) starting from the chain rule
\begin{align*}
\int d\xi''\,\frac{\delta J(\xi'')}{\delta\Psi(\xi)}\frac{\delta}{\delta J(\xi'')}=\frac{\delta}{\delta\Psi(\xi)},
\end{align*}
(ii) substituting Eq.\ (\ref{Gamma-deriv1}) with $I=0$ into its numerator $J(\xi'')$, 
(iii) operating the resulting expression to Eq.\ (\ref{dlnZ1}) with $\xi\rightarrow\xi'$ and $I=0$, (iv) setting $J=0$ subsequently, 
and (v) using Eqs.\ (\ref{dlnZ11}) and (\ref{Gamma^(n)-def}).

It is often convenient to regard $G^{-1}(\xi,\xi')\!=\!G_{jj'}^{-1}(x,x')$ as the $jj'$ element of the $2\!\times\!2$ matrix $\hat{G}^{-1}(x,x')$
in the particle-hole (i.e., Nambu) space.
Let us divide $\hat{G}^{-1}$ into the noninteracting part $\hat{G}_0^{-1}$ and the self-energy $\hat\Sigma$,
\begin{align}
\hat{G}^{-1}(x,x')=\hat{G}_0^{-1}(x,x')-\hat\Sigma(x,x'),
\label{DB}
\end{align}
which is equivalent to the Dyson-Beliaev equation, 
as seen by multiplying both sides by $\hat{G}(x',x'')$ from the right-hand side and integrating over $x'$.
It follows from Eq.\ (\ref{S_0}) that $\hat{G}_0^{-1}$ is given by
\begin{align}
\hat{G}_0^{-1}(x,x') = \left[i\hat\sigma_2\frac{\partial}{\partial\tau}-\hat\sigma_1\left(\frac{\hat{\bf p}^2}{2m} -\mu\right)\right]\delta(x-x') ,
\label{hatG_0^-1}
\end{align}
where $\hat\sigma_i$ $(i=1,2,3)$ is the $i$th Pauli matrix.

It should be noted that the present arrangement of $\hat{G}=(G_{jj'})$ in the Nambu space, 
which naturally results from Eq.\ (\ref{dlnZ11}),  differs from that of 
$\hat{G}_{\rm prev}$ used in the previous studies \cite{Kita09,Kita10,Kita14};
they are connected by $\hat{G}_{\rm prev}\!=\!\hat{G}(-i\hat\sigma_2)$.

\subsection{Goldstone's Theorem I}

Functional $\Gamma_{J}$ is invariant under the gauge transformation 
$(\Psi_{J1}(x),\Psi_{J2}(x))\rightarrow (\Psi_{J1}(x)e^{i\chi},\Psi_{J2}(x)e^{-i\chi})$,
where $\chi$ is a constant. 
Thus $\delta\Gamma_J/\delta\chi\!=\!0$ holds, which can be transformed by using Eq.\ (\ref{dPsi}) into
\begin{align*}
\sum_{j=1}^2\frac{\delta\Gamma_J}{\delta\Psi(\xi)}(-1)^{j-1} \left[\Psi(\xi)+\delta\Psi(\xi)\right]=0.
\end{align*}
Substituting Eq.\ (\ref{Gamma_J-exp}) and setting the coefficients of $n$th order equal to zero, 
we obtain
\begin{align}
&\,\left[(-1)^{j_1-1}+\cdots +(-1)^{j_n-1}\right]\Gamma^{(n)}(\xi_1,\cdots,\xi_n)
\notag \\
=&\,-\int d\xi \,\Gamma^{(n+1)}(\xi_1,\cdots,\xi_n,\xi)(-1)^{j-1}\Psi(\xi) .
\label{GoldstoneI}
\end{align}
Note that differentiation of Eq.\ (\ref{GoldstoneI}) with respect to $\Psi(\xi_{n+1})$ 
yields the $(n\!+\!1)$th identity by using Eq.\ (\ref{Gamma^(n)-def}).

The case of $n\!=\!1$ is expressible by substituting Eq.\ (\ref{Gamma^(1,2)}) and adopting
the vector-matrix notation of Eqs.\ (\ref{vecpsi})  and (\ref{DB}) as 
\begin{align}
\int dx' \hat\sigma_3\left[\hat{G}_{0}^{-1}(x,x')-\hat\Sigma(x,x')\right]\hat\sigma_3\vec\Psi(x')=\vec{0},
\label{HP}
\end{align}
which extends the Hugenholtz-Pines relation \cite{HP59} to inhomogeneous systems.
Next, we set $n\!=\!2$ in Eq.\ (\ref{GoldstoneI}),  substitute Eq.\ (\ref{Gamma^(2)}) with Eq.\ (\ref{DB}),
and use $[(-1)^{j-1}+(-1)^{j'-1}] G_0^{-1}(\xi,\xi')=0$ as seen from Eq.\ (\ref{hatG_0^-1}).
The procedure yields
\begin{align}
&\,\left[(-1)^{j-1}+(-1)^{j'-1}\right]\Sigma(\xi,\xi')
\notag \\
=&\,-\int d\xi_1 \,\Gamma^{(3)}(\xi,\xi',\xi_1)(-1)^{j_1-1}\Psi(\xi_1),
\label{n=2-identity}
\end{align}
which connects the anomalous self-energy $\Sigma_{jj}(x,x')$ with the three-point vertex.

The $n\!=\!1$ identity (\ref{HP}) has been presented as the key result 
from Goldstone's theorem I \cite{GSW62,Weinberg96,JL64,Swanson92}.
On the other hand, higher-order identities have turned out equally important. 
Among them, the $n\!=\!2$ identity was obtained by Gavoret and Nozi\`eres; see the second equality of Eq.\ (5.4). Later, it was used by Nepomnyashchi\u{i} and Nepomnyashchi\u{i} to show that the anomalous self-energy vanishes in the low energy-momentum limit \cite{Nepomnyashchii75,Nepomnyashchii78}. Castellani {\it et al}.\ \cite{CCPS97} derived and considered the identities of $n\!\leq\! 3$ in their renormalization-group study at $T=0$. 
The $n\!=\!2$ identity (\ref{n=2-identity}) will play a crucial role in the derivation of the two-particle Green's function below.

\subsection{Luttinger-Ward Functional}

Following Luttinger and Ward \cite{LW60},
we formally write $\Gamma$ in terms of another unknown functional $\Phi$ as \cite{Kita09,Kita14}
\begin{align}
\Gamma =&\, \frac{1}{2}\vec\Psi^{\rm T}\hat\sigma_3 \hat{G}_{0}^{-1}\hat\sigma_3\vec\Psi
\notag \\
&\,+\frac{1}{2}{\rm Tr} \,\bigl\{\ln \bigl[\bigl(-i\hat\sigma_2\bigr)\bigl(\hat{G}_{0}^{-1}-\hat\Sigma\bigr)\bigr]+\hat\Sigma\hat{G}\bigr\}
+\Phi,
\label{LW}
\end{align}
where $^{\rm T}$ denotes transpose, $\hat\Sigma=\hat\Sigma[\vec\Psi,\hat{G}]$ with $\hat{G}=\hat{G}[\vec\Psi]$, and integration over $\xi\equiv (j,x)$ is implied.
Then $\Gamma^{(1)}(\xi)=\delta\Gamma/\delta\Psi(\xi)$ acquires the expression
\begin{align}
\Gamma^{(1)}(\xi)=\int d\xi' G_0^{-1}(\xi,\xi')(-1)^{j+j'} \Psi(\xi')+\frac{\partial\Phi}{\partial\Psi(\xi)} ,
\label{Gamma^(1)}
\end{align}
where we have used Eq.\ (\ref{dGamma2}) to omit implicit dependences through $G$ in the differentiation;
see also the comment in the paragraph below Eq.\ (\ref{dGamma}) concerning the use of $\partial$ instead of $\delta$.
The right-hand side of Eq.\ (\ref{Gamma^(1)}) should be identical with the left-hand side of Eq.\ (\ref{HP}).
Thus, we obtain
\begin{subequations}
\label{dPhi}
\begin{align}
\frac{\partial\Phi}{\partial\Psi(\xi)}=&\, \int d\xi' \,\Sigma(\xi,\xi')(-1)^{j+j'-1}\Psi(\xi') .
\label{dPhi1}
\end{align}
Similarly, substitution of Eq.\ (\ref{LW}) into Eq.\ (\ref{dGamma2}) yields
\begin{align}
\frac{\partial\Phi}{\partial G(\xi',\xi)}=&\,-\frac{1}{2}\Sigma(\xi,\xi') ,
\label{dPhi2}
\end{align}
\end{subequations}
where we have used Eqs.\ (\ref{dGamma1}) and (\ref{DB}).
These are the two basic relations concerning $\Phi$. 

\section{Two-Particle Green's functions\label{sec3}}

We will derive expressions of two-particle Green's functions
based on the Dyson-Beliaev equation (\ref{DB}) and Hugenholtz-Pines relation (\ref{HP}).

\subsection{Variations $\delta\vec\Psi$ and $\delta\hat{G}$ under perturbation}

To this end, we switch on the infinitesimal perturbation given in terms of Eq.\ (\ref{Z})
by $(J,I)=(0,0)\rightarrow  (0,\delta I)$  once again \cite{Schwinger51,KB62}.
Accordingly, Eqs.\ (\ref{DB}) and (\ref{HP}), 
which are expressible concisely as $(\hat{G}_0^{-1}-\hat\Sigma)\hat{G}=\hat{1}$ and 
$\hat\sigma_3(\hat{G}_0^{-1}-\hat\Sigma)\hat\sigma_3\vec\Psi=\vec{0}$,  are modified into
\begin{subequations}
\label{DB-HP}
\begin{align}
\left(\hat{G}_0^{-1}-\hat\Sigma_I-\delta\hat{I}^{({\rm s})}\right)\hat{G}_I=&\,\hat{1},
\label{DB-HP1}
\\
\left[\hat\sigma_3\left(\hat{G}_0^{-1}-\hat\Sigma_I\right)\hat\sigma_3+\delta\hat{I}^{({\rm s})}\right]\vec\Psi_I=&\,\vec{0},
\label{DB-HP2}
\end{align}
\end{subequations}
where $\delta\hat{I}^{({\rm s})}$ is defined by
\begin{align}
\delta\hat{I}^{({\rm s})}(x,x')\equiv \delta\hat{I}(x,x')+\delta\hat{I}^{\,{\rm T}}(x',x).
\label{calU^(s)}
\end{align}
Equation (\ref{DB-HP}) together with $\hat\sigma_3\hat{G}_0^{-1}\hat\sigma_3=-\hat{G}_0^{-1}$ implies that the perturbation
gives rise to the direct variation $\hat{G}_0^{-1}\rightarrow \hat{G}_0^{-1}-\delta\hat{I}^{({\rm s})}$ 
plus the implicit one through the self-energies, the total of which 
cannot be described by the simple replacement $\hat{G}^{-1}\rightarrow \hat{G}^{-1}-\delta\hat{I}^{({\rm s})}$,
however. 
This point was overlooked in the previous study \cite{Kita10};
Eq.\ (\ref{DB-HP2}) forms one of the main corrections.

Let us collect terms of the first order in $\delta I$ from Eq.\ (\ref{DB-HP}).
The resulting equations can be written in terms of $\delta \hat{G}\equiv \hat{G}_I-\hat{G}$
and $\delta \hat\Sigma\equiv \hat\Sigma_I-\hat\Sigma$ as
\begin{subequations}
\label{dDB-HP1}
\begin{align}
\delta\hat{G}=&\,\hat{G}\left(\delta\hat{I}^{({\rm s})}+\delta\hat{\Sigma} \right)\hat{G},
\label{dDB-HP11}
\\
\hat{\sigma}_3\hat{G}^{-1}\hat{\sigma}_3\,\delta\vec{\Psi}=&\,\left(-\delta\hat{I}^{({\rm s})}+\hat{\sigma}_3\,\delta\hat{\Sigma}\,\hat{\sigma}_3\right) \vec{\Psi}.
\label{dDB-HP12}
\end{align}
\end{subequations}
Since $\hat{\Sigma}=\hat{\Sigma}[\Psi,G]$, moreover, we can express $\delta\Sigma$ as
\begin{align}
\delta\Sigma(\xi_1,\xi_1')=&\,\frac{\partial\Sigma(\xi_1,\xi_1')}{\partial G(\xi_2,\xi_2')}\delta G(\xi_2,\xi_2')
+\frac{\partial\Sigma(\xi_1,\xi_1')}{\partial \Psi(\xi_2)}\delta \Psi(\xi_2),
\label{dSigma}
\end{align}
where integration over repeated arguments is implied.
Substitution of Eq.\ (\ref{dPhi2}) into Eq.\ (\ref{dSigma}) yields 
\begin{align}
\delta\Sigma(\xi_1,\xi_1')=&\,-\frac{1}{2}\Gamma^{(4{\rm i})}(\xi_1,\xi_1';\xi_2,\xi_2')\delta G(\xi_2,\xi_2') 
\notag \\
&\,+\Gamma^{(3{\rm i}){\rm T}}(\xi_1,\xi_1';\xi_2)\delta \Psi(\xi_2) ,
\label{dSigma-Gamma}
\end{align}
where $\Gamma^{(4{\rm i})}$ is the irreducible four-point vertex defined by
\begin{subequations}
\begin{align}
\Gamma^{(4{\rm i})}(\xi_1,\xi_1';\xi_2,\xi_2')\equiv  &\,4\frac{\partial^2\Phi}{\partial G(\xi_1',\xi_1)\partial G(\xi_2,\xi_2')}
\notag \\
=&\,\Gamma^{(4{\rm i})}(\xi_2',\xi_2;\xi_1',\xi_1).
\label{Gamma^(4i)}
\end{align}
Similarly, we have introduced the irreducible three-point vertex by
$\Gamma^{(3{\rm i}){\rm T}}(\xi_1,\xi_1';\xi_2)\equiv \partial\Sigma(\xi_1,\xi_1')/\partial \Psi(\xi_2)$,
which can be transformed by using Eqs.\ (\ref{dPhi}) and (\ref{Gamma^(4i)}) into
\begin{align}
\Gamma^{(3{\rm i}){\rm T}}(\xi_1,\xi_1';\xi_2)=&\, \Gamma^{(4{\rm i})}(\xi_1,\xi_1';\xi_2',\xi_2) (-1)^{j_2+j_2'-1}\Psi(\xi_2')
\notag \\
=&\, \Psi(\xi_2')(-1)^{j_2+j_2'-1}\Gamma^{(4{\rm i})}(\xi_2,\xi_2';\xi_1',\xi_1) 
\notag \\
\equiv &\,\Gamma^{(3{\rm i})}(\xi_2;\xi_1',\xi_1).
\label{Gamma^(3i)-def}
\end{align}
\end{subequations}

To proceed further, we adopt the notation \cite{Kita10} 
\begin{subequations}
\label{VecMat}
\begin{align}
\langle\xi|\vec\Psi \equiv&\,\Psi(\xi),
\\
\langle\xi|\hat{\sigma}_3|\xi'\rangle \equiv&\,\delta(\xi,\xi')(-1)^{j-1},
\\
\langle\xi|\hat{G}|\xi'\rangle \equiv&\,G(\xi,\xi'),
\\
\langle \xi,\xi'|\vec{G}\equiv&\,G(\xi,\xi'),
\label{vecG}
\\
\langle \xi_1,\xi_1'|\underline{GG}|\xi_2,\xi_2'\rangle \equiv&\,G(\xi_1,\xi_2)G(\xi_1',\xi_2') ,
\\
\langle \xi_1,\xi_1'|\underline{1}|\xi_2,\xi_2'\rangle \equiv&\,\delta(\xi_1,\xi_2)\delta(\xi_1',\xi_2') ,
\\
\langle \xi_1|\underline{\Psi}^{(3)}|\xi_2,\xi_2'\rangle \equiv&\,\delta(\xi_1,\xi_2)\Psi(\xi_2')
\notag \\
\equiv&\, \langle \xi_2',\xi_2|\underline{\Psi}^{(3){\rm T}}|\xi_1\rangle,
\\
\langle \xi_1|\underline{\Psi}^{(3\sigma)}|\xi_2,\xi_2'\rangle \equiv&\,\delta(\xi_1,\xi_2)\Psi(\xi_2')(-1)^{j_2'-1}
\notag \\
\equiv&\, \langle \xi_2',\xi_2|\underline{\Psi}^{(3\sigma){\rm T}}|\xi_1\rangle,
\\
\langle \xi_1,\xi_1'|\underline\Gamma^{(4{\rm i})}|\xi_2,\xi_2'\rangle \equiv&\,\Gamma^{(4{\rm i})}(\xi_1,\xi_1';\xi_2,\xi_2'),
\\
\langle \xi_1|\underline\Gamma^{(3{\rm i})}|\xi_2,\xi_2'\rangle \equiv&\,\Gamma^{(3{\rm i})}(\xi_1;\xi_2,\xi_2')
\notag \\
\equiv&\, \langle \xi_2',\xi_2|\underline{\Gamma}^{(3{\rm i}){\rm T}}|\xi_1\rangle ,
\end{align}
\end{subequations}
together with $\delta\vec\Sigma$ and $\delta\vec{I}$ which are defined similarly as $\vec{G}$ in Eq.\ (\ref{vecG}).
Using the notation, we can express Eq.\ (\ref{Gamma^(3i)-def}) as
\begin{subequations}
\label{Gamma^i-connections}
\begin{align}
\underline\Gamma^{(3{\rm i})}=&\, -\hat\sigma_3\underline{\Psi}^{(3\sigma)}\underline\Gamma^{(4{\rm i})},
\\
\underline\Gamma^{(3{\rm i}){\rm T}}=&\, -\underline\Gamma^{(4{\rm i})}\underline{\Psi}^{(3\sigma){\rm T}}\hat\sigma_3,
\end{align}
\end{subequations}
and Eq.\ (\ref{dDB-HP1}) can be written alternatively as
\begin{subequations}
\label{dDB-HP2}
\begin{align}
\delta\vec{G}=&\,\underline{GG}\left(\delta\vec{I}^{\,({\rm s})}+\delta\vec{\Sigma} \right),
\label{dDB-HP21}
\\
\hat{\sigma}_3\hat{G}^{-1}\hat{\sigma}_3\,\delta\vec{\Psi}=&\,-\underline{\Psi}^{(3)}\delta\vec{I}^{\,({\rm s})}
+\hat{\sigma}_3\underline{\Psi}^{(3\sigma)}\delta\vec\Sigma.
\label{dDB-HP22}
\end{align}
\end{subequations}
For example, we can reproduce Eq.\ (\ref{dDB-HP11}) from Eq.\ (\ref{dDB-HP21}) by multiplying the latter by $\langle \xi_1,\xi_1'|$ from the left, inserting $|\xi_2,\xi_2'\rangle\langle \xi_2,\xi_2'|$ after $\underline{GG}$, and using $G(\xi_2,\xi_2')=G(\xi_2',\xi_2)$.
Similarly, Eq.\ (\ref{dSigma-Gamma}) reads
\begin{align}
\delta\vec\Sigma=-\frac{1}{2}\,\underline{\Gamma}^{(4{\rm i})}\delta\vec{G}+\underline{\Gamma}^{(3{\rm i}){\rm T}}\delta\vec\Psi.
\label{dSigma-Gamma2}
\end{align}

The coupled equations (\ref{dDB-HP2}) and (\ref{dSigma-Gamma2}) are solved as follows.
First, after substituting Eq.\ (\ref{dSigma-Gamma2}), we can solve Eq.\ (\ref{dDB-HP21}) formally in terms of $\delta\vec{G}$ as
\begin{align}
\delta \vec{G}= &\, \underline{g}^{(4)} \left(\delta\vec{I}^{\,({\rm s})} +\underline{\Gamma}^{(3{\rm i}){\rm T}}\delta \vec\Psi\right) ,
\label{dG-sol1}
\end{align}
with 
\begin{align}
\underline{g}^{(4)}\equiv \left(\underline{1}+\frac{1}{2}\,\underline{GG}\,\underline{\Gamma}^{(4{\rm i})}\right)^{-1}\underline{GG} .
\label{g^(4)}
\end{align}
It is useful at this stage to introduce the full four- and three-point vertices by
\begin{subequations}
\label{Gamma^(4,3)}
\begin{align}
\underline{\Gamma}^{(4)}\equiv &\,\left(\underline{1}+\frac{1}{2}\,\underline{\Gamma}^{(4{\rm i})}\underline{GG}\right)^{-1}\underline{\Gamma}^{(4{\rm i})},
\label{Gamma^(4,3)-1}
\\
\underline{\Gamma}^{(3)}\equiv&\,-\hat\sigma_3\underline{\Psi}^{(3\sigma)}\underline{\Gamma}^{(4)},
\label{Gamma^(4,3)-2}
\\
\underline{\Gamma}^{(3){\rm T}}\equiv &\,-\underline{\Gamma}^{(4)}\underline{\Psi}^{(3\sigma){\rm T}}\hat\sigma_3 .
\label{Gamma^(4,3)-3}
\end{align}
\end{subequations}
Indeed, Eq.\ (\ref{dG-sol1}) is expressible alternatively in terms of the vertices as
\begin{align}
\delta\vec{G}=&\,\underline{GG}\left(\underline{1}-\frac{1}{2}\,\underline{\Gamma}^{(4)}\underline{GG}\right) \delta\vec{I}^{\,({\rm s})} 
+\underline{GG}\,\underline{\Gamma}^{(3){\rm T}}\delta\vec\Psi,
\label{dG-sol2}
\end{align}
where we have used the matrix identity $(\underline{1}+\underline{A}\,\underline{B})^{-1}\underline{A}=\underline{A}(\underline{1}+\underline{B}\,\underline{A})^{-1}=\underline{A}-\underline{A}(\underline{1}+\underline{B}\,\underline{A})^{-1}\underline{B}\,\underline{A}$
and Eq.\ (\ref{Gamma^i-connections}).
With $\delta\vec{G}\!=\!-\underline{GG}\,\delta\vec{G}^{-1}$ which is equivalent to $\delta\hat{G}\!=\!-\hat{G}\,\delta\hat{G}^{-1}\hat{G}$, we can
transform Eq.\ (\ref{dG-sol2}) further into
\begin{align}
\delta\vec{G}^{-1}=&\,-\left(\underline{1}-\frac{1}{2}\,\underline{\Gamma}^{(4)}\underline{GG}\right) \delta\vec{I}^{\,({\rm s})} 
-\underline{\Gamma}^{(3){\rm T}}\delta\vec\Psi.
\label{dG-sol3}
\end{align}
Noting Eq.\ (\ref{Gamma^(2)}), we can identify $\underline{\Gamma}^{(3){\rm T}}$ above as $\Gamma^{(3)}$ of Eq.\ (\ref{Gamma^(n)-def}) defined for $\delta I\!=\!0$.
Thus, we conclude that ${\Gamma}^{(3){\rm T}}(\xi_1,\xi_1';\xi_2)\equiv \langle \xi_1,\xi_1'|\underline{\Gamma}^{(3){\rm T}}|\xi_2\rangle$ is symmetric with respect to any permutation of its arguments, and
\begin{align}
\underline{\Gamma}^{(3){\rm T}}=\underline{\Gamma}^{(3)}
\label{Gamma^(3)T=Gamma^(3)}
\end{align}
holds.
This observation will play a crucial role below.
It should also be noted that the vertices may acquire asymmetry
in practical studies of using approximate $\Phi$ in Eq.\ (\ref{Gamma^(4i)}).
With this possibility in mind, we will proceed with keeping the formal distinction between $\underline{\Gamma}^{(3)}$ and $\underline{\Gamma}^{(3){\rm T}}$.

Next, we focus on Eq.\ (\ref{dDB-HP22}) and substitute Eq.\ (\ref{dSigma-Gamma2}) with Eq.\  (\ref{dG-sol1}) into it.
We then obtain
\begin{align}
&\,\biggl[\hat{\sigma}_3\hat{G}^{-1}\hat\sigma_3 -\hat\sigma_3\underline{\Psi}^{(3\sigma)}\biggl(\underline{1} -\frac{1}{2}\,\underline{\Gamma}^{(4{\rm i})}\underline{g}^{(4)}\biggr)\underline{\Gamma}^{(3{\rm i}){\rm T}}\biggr]
\delta\vec{\Psi}
\notag \\
=&\,-\biggl(\underline{\Psi}^{(3)}
+\frac{1}{2}\,\hat\sigma_3\underline{\Psi}^{(3\sigma)}\underline{\Gamma}^{(4{\rm i})}\underline{g}^{(4)} \biggr)\delta\vec{I}^{\,({\rm s})} .
\label{dPsi-eq}
\end{align}
The prefactor of $\delta\vec{\Psi}$ can be transformed as
\begin{align}
&\,\hat{\sigma}_3\hat{G}^{-1}\hat\sigma_3 -\hat\sigma_3\underline{\Psi}^{(3\sigma)}\biggl(\underline{1}-\frac{1}{2}\,\underline{\Gamma}^{(4{\rm i})}\underline{g}^{(4)}\biggr)
\underline{\Gamma}^{(3{\rm i}){\rm T}}
\notag \\
=&\, \hat{\sigma}_3\hat{G}_0^{-1}\hat\sigma_3-\hat{\sigma}_3\hat\Sigma\hat\sigma_3
-\hat\sigma_3\underline{\Psi}^{(3\sigma)}\biggl(\underline{1}+\frac{1}{2}\,\underline{\Gamma}^{(4{\rm i})}\underline{GG}\biggr)^{\!\!-1}
\underline{\Gamma}^{(3{\rm i}){\rm T}}
\notag \\
=&\, -\hat{G}_0^{-1}-\hat{\sigma}_3\hat\Sigma\hat\sigma_3
-\hat\sigma_3\underline{\Psi}^{(3\sigma)}
\underline{\Gamma}^{(3){\rm T}}
\notag \\
=&\, -\hat{G}_0^{-1}-\hat{\sigma}_3\hat\Sigma\hat\sigma_3
-\hat\sigma_3\bigl(-\hat\sigma_3\hat\Sigma-\hat\Sigma\hat\sigma_3\bigr)
\notag \\
=&\, -\hat{G}^{-1} ,
\label{prefactor}
\end{align}
where we have successively used Eqs.\ (\ref{DB}), 
(\ref{g^(4)}), (\ref{hatG_0^-1}), (\ref{Gamma^(4,3)}), (\ref{Gamma^(3)T=Gamma^(3)}), and (\ref{n=2-identity}). 
Thus, the prefactor has been identified as $-\hat{G}^{-1}$,
due mainly to the $n\!=\!2$ identity (\ref{n=2-identity}) which has been incorporated additionally
in the present study.
Equation (\ref{prefactor}) forms the second correction to the previous study \cite{Kita10}.
Indeed, the result will lead us to the conclusion that the single- and two-particle Green's functions share poles,
in agreement with the Gavoret-Nozi\`eres theory \cite{GN64}.

Let us substitute Eq.\ (\ref{prefactor}) into Eq.\ (\ref{dPsi-eq}) and write $\hat\sigma_3\underline{\Psi}^{(3\sigma)}\underline{\Gamma}^{(4{\rm i})}\underline{g}^{(4)} =-\underline{\Gamma}^{(3)}\underline{GG}$ based on Eqs.\ (\ref{Gamma^i-connections}), (\ref{g^(4)}), and (\ref{Gamma^(4,3)}).
We thereby obtain
\begin{subequations}
\label{dPsidG-sol}
\begin{align}
\delta\vec{\Psi}
=\hat{G}\biggl(\underline{\Psi}^{(3)}
-\frac{1}{2}\,\underline{\Gamma}^{(3)}\underline{GG} \biggr)\delta \vec{I}^{\,({\rm s})}  ,
\label{dPsi-sol}
\end{align}
so that Eq.\ (\ref{dG-sol2}) now reads
\begin{align}
\delta\vec{G}=&\,\left[\underline{GG}\left(\underline{1}-\frac{1}{2}\,\underline{\Gamma}^{(4)}\underline{GG}\right) 
\right.
\notag \\
&\,\left. +\,\underline{GG}\,\underline{\Gamma}^{(3){\rm T}}\,\hat{G}\biggl(\underline{\Psi}^{(3)}
-\frac{1}{2}\,\underline{\Gamma}^{(3)}\underline{GG} \biggr)\right]\delta \vec{I}^{\,({\rm s})} .
\label{dG-sol}
\end{align}
\end{subequations}
It is worth noting that $\hat{G}\underline{\Psi}^{(3)}\delta \vec{I}^{\,({\rm s})}$ in Eq.\ (\ref{dPsi-sol}) is also
expressible as
\begin{align}
\hat{G}\,\underline{\Psi}^{(3)}\delta \vec{I}^{\,({\rm s})} =\hat{G}\,\delta \hat{I}^{\,({\rm s})} \vec\Psi .
\label{GI^(s)Psi}
\end{align}
The equivalence can be seen easily by operating $\langle \xi|$ from the left,
inserting $|\xi_1\rangle\langle\xi_1|$ or $|\xi_1\xi_1'\rangle\langle\xi_1\xi_1'|$ appropriately,
and using Eq.\ (\ref{VecMat}).

\subsection{Expressions of $G^{(3)}$ and $G^{(4)}$}

Let us define the $n$-point Green's function in terms of Eq.\ (\ref{Z}) with $I=0$ by
\begin{align}
G^{(n)}(\xi_1,\cdots,\xi_n)=(-1)^{n-\left[\frac{n}{2}\right]}\frac{\delta^n \ln Z_J}{\delta J(\xi_1)\cdots J(\xi_n)}\Biggr|_{J=0},
\label{G^(n)-def}
\end{align}
where $\left[\frac{n}{2}\right]$ denotes the largest integer that does not exceed $\frac{n}{2}$.
Equation (\ref{G^(n)-def}) is the $n$th cumulant composed of connected Feynman diagrams.
The first two of them are given in terms of the functions in Eq.\ (\ref{dlnZ}) by $G^{(1)}(\xi_1)\!=\!\Psi(\xi_1)$
and $G^{(2)}(\xi_1,\xi_2)\!=\!G(\xi_1,\xi_2)$.
Noting $G^{(n+1)}(\xi_1,\cdots,\xi_{n+1})\propto\delta G^{(n)}(\xi_1,\cdots,\xi_{n})/\delta J(\xi_{n+1})$,
we can obtain $G^{(3)}$ and $G^{(4)}$ successively from Eq.\ (\ref{dlnZ11}).
They are expressible concisely in terms of 
\begin{align}
{\cal G}^{(n)}(\xi_1,\cdots,\xi_n)\equiv (-1)^{\left[\frac{n}{2}\right]}\langle T_\tau \psi(\xi_1)\cdots\psi(\xi_n)\rangle .
\label{calG^(n)}
\end{align}
with abbreviating $G^{(4)}(\xi_1,\xi_2,\xi_3,\xi_4)\!\equiv\!G^{(4)}_{1234}$, etc., as
\begin{subequations}
\label{G^(3,4)-1}
\begin{align}
G^{(3)}_{123}
=&\,{\cal G}^{(3)}_{123} -\Psi_1G_{23}
-\Psi_2 G_{31}
-\Psi_3G_{12}
+\Psi_1\Psi_2\Psi_3 ,
\label{G^(3,4)-11}
\end{align}
\begin{align}
G^{(4)}_{1234}
=&\,{\cal G}^{(4)}_{1234} +\Psi_1G^{(3)}_{234}
+\Psi_2G^{(3)}_{341}+\Psi_3G^{(3)}_{412}
+\Psi_4 G^{(3)}_{123}
\notag \\
&\, -G_{12}G_{34}-G_{13}G_{24}-G_{14}G_{23}
\notag \\
&\,+\Psi_1\Psi_2G_{34}+\Psi_1\Psi_3G_{24}
+\Psi_1\Psi_4G_{23}
\notag \\
&\,
+\Psi_2\Psi_3G_{14}+\Psi_2\Psi_4G_{13}+\Psi_3\Psi_4G_{23}
\notag \\
&\,
-\Psi_1\Psi_2\Psi_3\Psi_4.
\label{G^(3,4)-12}
\end{align}
\end{subequations}
The correctness of Eqs.\ (\ref{G^(3,4)-11}) and (\ref{G^(3,4)-12}) can be seen in the fact that
terms with $\Psi$ and $G$ on their right-hand sides appropriately
remove all the disconnected contributions from ${\cal G}^{(n)}$ for $n=3,4$.

On the other hand, 
one can show using Eqs.\ (\ref{Z}) and (\ref{dlnZ}) that Eq.\ (\ref{G^(3,4)-1}) can be written alternatively
in terms of the variations of $\vec\Psi_I$ and $G_I$ with respect to $I$ by
\begin{subequations}
\label{G^(3,4)-2}
\begin{align}
G^{(3)}_{123}=\frac{\delta\Psi_1}{\delta I_{32}}\Biggr|_{I=0}
-\Psi_2G_{13}-\Psi_3G_{12} ,
\label{G^(3,4)-21}
\end{align}
\begin{align}
G^{(4)}_{1234}
=&\,\frac{\delta G_{12}}{\delta I_{43}}\Biggr|_{I=0}+\Psi_3G^{(3)}_{124}
+\Psi_4G^{(3)}_{123}
\notag \\
&\, -G_{13}G_{24}-G_{14}G_{23} .
\label{G^(3,4)-22}
\end{align}
\end{subequations}

Let us express $\delta\Psi(\xi_1)=\langle \xi_1|\delta\vec\Psi$  in Eq.\ (\ref{G^(3,4)-21}),
substitute Eq.\ (\ref{dPsi-sol}) with Eq.\ (\ref{calU^(s)}), and perform the differentiation.
Noting that $\Gamma^{(3)}$ and $G$ are symmetric with respect to the arguments,
we obtain
\begin{subequations}
\label{G^(3,4)}
\begin{align}
&\,G^{(3)}(\xi_1,\xi_2,\xi_3)
\notag \\
=&\,-G(\xi_1,\xi_1')G(\xi_2,\xi_2')G(\xi_3,\xi_3')\Gamma^{(3)}(\xi_1';\xi_2',\xi_3') .
\label{G^(3)}
\end{align}
We also write  $\delta G(\xi_1,\xi_2)=\langle \xi_1,\xi_2|\delta\vec{G}$   in Eq.\ (\ref{G^(3,4)-22}),
substitute Eq.\ (\ref{dG-sol}), perform the differentiation, 
and use Eq.\ (\ref{G^(3)}).
The procedure yields
\begin{align}
&\,G^{(4)}(\xi_1,\xi_2,\xi_3,\xi_4)
\notag \\
=&\,-G(\xi_1,\xi_1')G(\xi_2,\xi_2')G(\xi_3,\xi_3')G(\xi_4,\xi_4')
\notag \\
&\,\times \left[\Gamma^{(4)}(\xi_1',\xi_2';\xi_3',\xi_4')
+\Gamma^{(3){\rm T}}(\xi_1',\xi_2';\xi_5) G(\xi_5,\xi_5')
\right.
\notag \\
&\,
\left.
\times\Gamma^{(3)}(\xi_5';\xi_3',\xi_4')
\right] ,
\label{G^(4)}
\end{align}
\end{subequations}
where $\Gamma^{(4)}$, $\Gamma^{(3){\rm T}}$, and  $\Gamma^{(3)}$ are given in  Eqs.\ (\ref{Gamma^(4,3)}). 

Functions $G^{(3)}$ and $G^{(4)}$ in Eq.\ (\ref{G^(3,4)}) are both connected, as they should,
and Eq.\ (\ref{G^(4)}) tells us clearly that $G^{(4)}$ shares poles with $G$, in agreement with the result of Gavoret and Nozi\`eres \cite{GN64}.
They should be symmetric with respect to any permutation of its arguments in the exact treatment;
the apparent asymmetry in Eq.\ (\ref{G^(3,4)})
originates from that of the irreducible vertex $\Gamma^{(4{\rm i})}$ defined by Eq.\ (\ref{Gamma^(4i)}) with which we have constructed $\Gamma^{(4)}$.
It should also be noted that both $G^{(3)}$ and $G^{(4)}$ will acquire asymmetry 
in practical studies of using some approximate $\Phi$ in Eq.\ (\ref{Gamma^(4i)}).

\section{Self-Energies and Condensate Wave Function\label{sec4}}

In this section, we derive (i) expressions of the self-energies in the Dyson-Beliaev equation
and (ii) the equation for the condensate wave function, i.e., the generalized Gross-Pitaevski\u{i} equation,
both in terms of $(G,\Gamma^{(3)},\Gamma^{(4)})$.
Subsequently, we will see that 
conservation laws are satisfied by the Dyson-Beliaev and Gross-Pitaevski\u{i} equations.

\subsection{Expressions of self-energies}

The Heisenberg equation of motion for the field operator $\hat\psi_{j}(x)\equiv e^{\tau \hat{H}}\hat\psi_{j}({\bf r})e^{-\tau \hat{H}}$ 
corresponding to Eq.\ (\ref{S}) is given by \cite{KB62}
\begin{align}
&\,\left[(-1)^{j}\frac{\partial}{\partial\tau}-\frac{\hat{\bf p}^2}{2m}+\mu\right]\hat\psi_{j}(x)
\notag \\
&\,-\int dx_1 U(x,x_1)\hat\psi_2(x_1)\hat\psi_{j}(x)\hat\psi_1(x_1) =0.
\label{EqMotion}
\end{align}
Taking its thermodynamic average yields
\begin{subequations}
\label{PsiG-EqMotion}
\begin{align}
&\,\left[(-1)^{j}\frac{\partial}{\partial\tau}-\frac{\hat{\bf p}^2}{2m}+\mu\right]\Psi_{j}(x)
\notag \\
&\,+\int dx_1 U(x,x_1){\cal G}^{(3)}_{2j1}(x_1,x,x_1) =0,
\label{Psi-EqMotion}
\end{align}
where ${\cal G}^{(3)}$ is defined by Eq.\ (\ref{calG^(n)}).
One can also show based on Eq.\ (\ref{EqMotion}) that ${\cal G}\equiv {\cal G}^{(2)}$
obeys \cite{KB62}
\begin{align}
&\,\left[(-1)^{j}\frac{\partial}{\partial\tau}-\frac{\hat{\bf p}^2}{2m}+\mu\right]{\cal G}_{jj'}(x,x')
\notag \\
&\,+\int dx_1 U(x,x_1){\cal G}^{(4)}_{2j1j'}(x_1,x,x_1,x')
=\delta_{j,3-j'}\delta(x,x') .
\label{calG-EqMotion}
\end{align}
\end{subequations}
We can construct the equation of motion for $G_{jj'}(x,x')\!=\!{\cal G}_{jj'}(x,x')\!+\!\Psi_j(x)\Psi_j(x')$
from Eq.\ (\ref{PsiG-EqMotion}).
It is given with $j\rightarrow 3-j$ by
\begin{align}
&\,\left[(-1)^{3-j}\frac{\partial}{\partial\tau}-\frac{\hat{\bf p}^2}{2m}+\mu\right]G_{3-j,j'}(x,x')
\notag \\
&\,+\int dx_1 U(x,x_1) \left[{\cal G}^{(4)}_{2,3-j,1,j'}(x_1,x,x_1,x')\right.
\notag \\
&\,\left.
+\,{\cal G}^{(3)}_{2,3-j,1}(x_1,x,x_1)\Psi_{j'}(x') \right]
=\delta_{jj'}\delta(x,x') .
\label{G-EqMotion}
\end{align}
Equation (\ref{G-EqMotion}) should be identical with Eq.\ (\ref{DB}) that can be written as $(\hat{G}_0^{-1}-\hat\Sigma)\hat{G}=\hat{1}$ in terms of $\hat{G}_0^{-1}$ in Eq.\ (\ref{hatG_0^-1}). 
Hence, we obtain
\begin{align}
&\,\Sigma(\xi,\xi')
\notag \\
=&-\!\frac{\delta}{\delta G(\xi',\xi'')}
\int\!  dx_1 U(x,x_1) 
\!\left[{\cal G}^{(4)}_{2,3-j,1,j''}(x_1,x,x_1,x'')\right.
\notag \\
&\,\left.
+{\cal G}^{(3)}_{2,3-j,1}(x_1,x,x_1)\Psi_{j''}(x'') \right].
\label{Sigma-1}
\end{align}
The terms in the square brackets of Eq.\ (\ref{Sigma-1}) can be transformed by using Eq.\ (\ref{G^(3,4)-1}) into
\begin{align}
 {\cal G}^{(4)}_{1234}+{\cal G}^{(3)}_{123}\Psi_4
=&\, G^{(4)}_{1234}
-\Psi_1 G^{(3)}_{234}
-\Psi_2 G^{(3)}_{134}
-\Psi_3 G^{(3)}_{124}
\notag \\
&\, +G_{12}G_{34}+G_{13}G_{24}
+G_{14}G_{23}
\notag \\
&\,
-\Psi_1\Psi_2 G_{34}
-\Psi_1\Psi_3 G_{24}
-\Psi_2\Psi_3 G_{14} .
\label{calG^4+calG^3Psi}
\end{align}
We use Eq.\ (\ref{calG^4+calG^3Psi}) in Eq.\ (\ref{Sigma-1}), substitute Eq.\ (\ref{G^(3,4)}),
perform the differentiation, 
and symmetrize the expression so that $\Sigma(\xi_1,\xi_1')\!=\! \Sigma(\xi_1',\xi_1)$ can be seen manifestly.
We thereby obtain
\begin{align}
&\,\Sigma(\xi,\xi')
\notag \\
=&\, \delta_{j,3-j'}\delta(x,x') U(x,x_1)
\bigl[\Psi_2(x_1)\Psi(x_1)\!-\!G_{21}(x_1,x_1)\bigr]
\notag \\
&\, +U(x,x')\bigl[\Psi_{3-j}(x)\Psi_{3-j'}(x')-G_{3-j,3-j'}(x,x')\bigr]
\notag \\
&\, +\frac{1}{2} U(x,x_1)\Bigl\{G_{3-j,j_2}(x,x_2) G_{2j_3}(x_1,x_3)G_{1j_3'}(x_1,x_3')
\notag \\
&\, \times \bigl[\Gamma^{(4)}(\xi_2,\xi';\xi_3,\xi_3')+\Gamma^{(3){\rm T}}(\xi_2,\xi';\xi_4)G(\xi_4,\xi_4')
\notag \\
&\,\times \Gamma^{(3)}(\xi_4';\xi_3,\xi_3')\bigr]
\notag \\
&\,-G_{3-j,j_2}(x,x_2)\bigl[
G_{2j_3}(x_1,x_3)\Psi_{1}(x_1)
\notag \\
&\,\hspace{25mm}+\Psi_{2}(x_1)G_{1j_3}(x_1,x_3)\bigr]\Gamma^{(3){\rm T}}(\xi_2,\xi';\xi_3)
\notag \\
&\, -\Psi_{3-j}(x)G_{2j_3}(x_1,x_3)G_{1j_3'}(x_1,x_3')\Gamma^{(3)}(\xi';\xi_3,\xi_3') 
\notag \\
&\, +(\xi\leftrightarrow\xi')\Bigr\},
\label{Sigma-Gamma}
\end{align}
where integration over repeated arguments is implied, and $(\xi\leftrightarrow\xi')$ denotes terms obtained from
the preceding three terms in the curly brackets by exchanging $\xi$ and $\xi'$.
The first two terms on the right-hand side are the Hartree and Fock terms that are expressible as Fig.\ \ref{fig:1}(a)-(d), 
whereas the third one represents correlation effects given diagrammatically by Fig.\ \ref{fig:1}(e)-(h).

It should be noted that there is arbitrariness in expressing the correlation term of Eq.\ (\ref{Sigma-Gamma}) in terms of $\Gamma^{(4)}$
and $\Gamma^{(3)}$, which are symmetric in the exact theory but may acquire asymmetry in approximate treatments.
We have removed it here so that 
the two Green's functions entering and leaving $x_1$ of the bare interaction vertex $U(x,x_1)$ in Eq.\ (\ref{Sigma-Gamma})
are linked with the latter two arguments of $\Gamma^{(4)}$, i.e., $(\xi_3,\xi_3')$.
The advantage of this choice is that the density fluctuation mode is naturally 
incorporated in $\Gamma^{(4)}$ even in approximate treatments.

Using Eq.\ (\ref{n=2-identity}) and following the argument of Nepomnyashchi\u{i} and Nepomnyashchi\u{i} \cite{Nepomnyashchii75,Nepomnyashchii78},
one can confirm oneself that diagrams (g) and (h) in Fig.\ \ref{fig:1} make the anomalous self-energy vanish in the low energy-momentum limit for homogeneous systems. Thus, the Nepomnyashchi\u{i} identity is naturally satisfied in our formulation
to remove the infrared divergence, thereby making practical calculations possible.

\subsection{Equation for the condensate wave function}

Let us express ${\cal G}^{(3)}$ in Eq.\ (\ref{Psi-EqMotion}) in terms of $(G^{(3)},G,\Psi)$
by using Eq.\ (\ref{G^(3,4)-11}) and substitute Eq.\ (\ref{G^(3)}) subsequently. 
We then obtain
\begin{align}
\left[(-1)^{j}\frac{\partial}{\partial\tau}-\frac{\hat{\bf p}^2}{2m}+\mu\right]\Psi_{j}(x)=\eta_{j}(x) ,
\label{GP-Eq}
\end{align}
with
\begin{align}
\eta_{j}(x)\equiv &\,U(x,x_1)\bigl[ G_{jj_2}(x,x_2)G_{2j_3}(x_1,x_3)G_{1j_3'}(x_1,x_3')
\notag \\
&\,\times \Gamma^{(3)}(x_2;x_3,x_3')-\Psi_{j}(x)G_{21}(x_1,x_1)
\notag \\
&\, -\Psi_{2}(x_1)G_{j1}(x,x_1)-\Psi_{1}(x_1)G_{2j}(x_1,x)
\notag \\
&\, +\Psi_{j}(x)\Psi_{2}(x_1)\Psi_{1}(x_1)\bigr],
\label{eta-def}
\end{align}
where integrations over $(x_1,\xi_2, \xi_2',\xi_3,\xi_3')$ are implied.
Equation (\ref{GP-Eq}) generalizes the Gross-Pitaevski\u{i} equation \cite{Gross61,Pitaevskii61} so as to incorporate the quasiparticle contribution and correlation effects in $\eta_j(x)$. 
It is equivalent to Eq.\ (\ref{HP}), i.e.,
the generalized Hugenholtz-Pines relation, in the exact theory.
However, they will be different in approximate treatments. 
We prefer Eq.\ (\ref{GP-Eq}) to Eq.\ (\ref{HP}), because conservation laws
are satisfied as seen below. 
Adopting Eq.\ (\ref{GP-Eq}), we should determine the chemical potential
so as to reproduce a branch of gapless excitations
in the single-particle channel.

\subsection{Conservation Laws}

We follow the argument of Kadanoff and Baym \cite{KB62,Baym61} 
to confirm that the number-, momentum-, and energy-conservation laws are satisfied 
in our formulation. 

First, Eq.\ (\ref{Sigma-Gamma}) satisfies $\Sigma(\xi,\xi')\!=\!\Sigma(\xi',\xi)$ and so do Green's functions
determined by the Dyson-Beliaev equation (\ref{DB}). 
Hence, criterion A of Kadanoff and Baym $(\hat{G}_0^{-1}\!-\!\hat\Sigma)\hat{G}\!=\!\hat{G}(\hat{G}_0^{-1}\!-\!\hat\Sigma)\!=\!\hat{1}$ is met,  thereby ensuring the number-conservation law.
Second, we consider their criterion B on the momentum- and energy-conservation laws.
The  Dyson-Beliaev equation (\ref{DB}) with Eq.\ (\ref{Sigma-Gamma}) can be written alternatively as
Eq.\ (\ref{G-EqMotion}). Moreover, Eq.\ (\ref{GP-Eq}) is equivalent to Eq.\ (\ref{Psi-EqMotion}).
Hence, our ${\cal G}(\xi,\xi')\!=\!G(\xi,\xi')\!-\!\Psi(\xi)\Psi(\xi')$ obeys Eq.\ (\ref{calG-EqMotion}),
where criterion B of Kadanoff and Baym ${\cal G}^{(4)}_{2211}(x_1,x,x_1,x)={\cal G}^{(4)}_{2211}(x,x_1,x,x_1)$ holds manifestly. Thus, the momentum- and energy-conservation laws are also fulfilled.

\section{System of Equations\label{sec5}}

\subsection{Derived equations}

Let us summarize our system of equations for easy reference. 
First, Green's functions $G(\xi,\xi')$ and the condensate wave function $\Psi(\xi)$ obey
the Dyson-Beliaev equation and generalized Gross-Pitaevski\u{i} equation given by
\begin{subequations}
\label{DB-GP}
\begin{align}
\left[G_0^{-1}(\xi,\xi_1)-\Sigma(\xi,\xi_1)\right] G(\xi_1,\xi')=&\, \delta(\xi,\xi'),
\\
\left[(-1)^{j}\frac{\partial}{\partial\tau}-\frac{\hat{\bf p}^2}{2m}+\mu\right]\Psi(\xi)=&\,\eta(\xi) ,
\end{align}
\end{subequations}
respectively, where $G_0^{-1}$ is defined by Eq.\ (\ref{hatG_0^-1}), $\xi\equiv (j,x)$ with $j=1,2$ and $x=({\bf r},\tau)$,
and integration over the repeated argument $\xi_1$ is implied.

\begin{figure}[b]
        \begin{center}
                \includegraphics[width=0.9\linewidth]{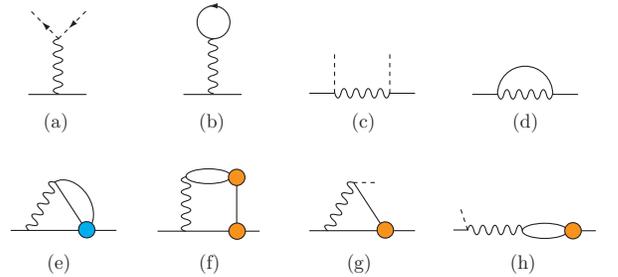}
        \end{center}
        \caption{Diagramatic expression of the self-energy. A wavy and dotted lines denote $U$ and $\Psi_j$, respectively.
        }
        \label{fig:1}
\end{figure}

Second, the self-energy $\Sigma$ and source function $\eta$ in Eq.\ (\ref{DB-GP}) 
are given in terms of $(\Psi,G,\Gamma^{(3)},\Gamma^{(4)})$ by
\begin{subequations}
\label{Sigma-eta}
\begin{align}
&\,\Sigma(\xi,\xi')
\notag \\
=&\, \delta_{j,3-j'}\delta(x,x') U(x,x_1)
\bigl[\Psi_2(x_1)\Psi(x_1)\!-\!G_{21}(x_1,x_1)\bigr]
\notag \\
&\, +U(x,x')\bigl[\Psi_{3-j}(x)\Psi_{3-j'}(x')-G_{3-j,3-j'}(x,x')\bigr]
\notag \\
&\, +\frac{1}{2} U(x,x_1)\Bigl\{G_{3-j,j_2}(x,x_2) G_{2j_3}(x_1,x_3)G_{1j_3'}(x_1,x_3')
\notag \\
&\, \times \bigl[\Gamma^{(4)}(\xi_2,\xi';\xi_3,\xi_3')+\Gamma^{(3){\rm T}}(\xi_2,\xi';\xi_4)G(\xi_4,\xi_4')
\notag \\
&\,\times \Gamma^{(3)}(\xi_4';\xi_3,\xi_3')\bigr]
\notag \\
&\,-G_{3-j,j_2}(x,x_2)\bigl[
G_{2j_3}(x_1,x_3)\Psi_{1}(x_1)
\notag \\
&\,+\Psi_{2}(x_1)G_{1j_3}(x_1,x_3)\bigr]\Gamma^{(3){\rm T}}(\xi_2,\xi';\xi_3)
\notag \\
&\, -\Psi_{3-j}(x)G_{2j_3}(x_1,x_3)G_{1j_3'}(x_1,x_3')\Gamma^{(3)}(\xi';\xi_3,\xi_3') 
\notag \\
&\, +(\xi\leftrightarrow\xi')\Bigr\},
\label{Sigma-Gamma2}
\end{align}
\begin{align}
\eta(\xi)\equiv &\,U(x,x_1)\bigl[ G_{jj_2}(x,x_2)G_{2j_3}(x_1,x_3)G_{1j_3'}(x_1,x_3')
\notag \\
&\,\times \Gamma^{(3)}(x_2;x_3,x_3')-\Psi_{j}(x)G_{21}(x_1,x_1)
\notag \\
&\, -\Psi_{2}(x_1)G_{j1}(x,x_1)-\Psi_{1}(x_1)G_{2j}(x_1,x)
\notag \\
&\, +\Psi_{j}(x)\Psi_{2}(x_1)\Psi_{1}(x_1)\bigr].
\label{eta-def2}
\end{align}
\end{subequations}
Equation (\ref{Sigma-Gamma2}) is expressible diagrammatically as Fig.\ \ref{fig:1}.

\begin{figure}[t]
        \begin{center}
                \includegraphics[width=0.8\linewidth]{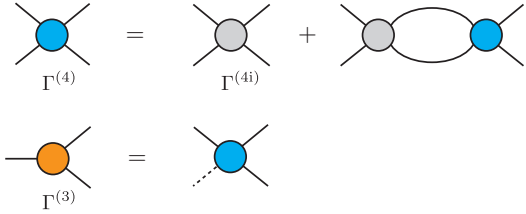}
        \end{center}
        \caption{Diagramatic expressions of the four- and three-point vertices. The dotted line denotes $\Psi$.}
        \label{fig:2}
\end{figure}

Third, the vertices $\underline{\Gamma}^{(4)}$ and $\underline{\Gamma}^{(3)}$ are defined by Eqs.\ (\ref{Gamma^(4,3)-1}) and (\ref{Gamma^(4,3)-2}), 
i.e., $\Gamma^{(4)}$ obeys
\begin{subequations}
\label{Gamma^(3,4)-2}
\begin{align}
\underline{\Gamma}^{(4)}=\underline{\Gamma}^{(4{\rm i})}-\frac{1}{2}\,\underline{\Gamma}^{(4{\rm i})}\underline{GG}\,\underline{\Gamma}^{(4)} ,
\label{Gamma^(4)-eq}
\end{align}
and $\Gamma^{(3)}(\xi_1;\xi_2,\xi_2')\equiv \langle\xi_1|\underline{\Gamma}^{(3)}|\xi_2,\xi_2'\rangle$ is given in terms of 
$\Gamma^{(4)}(\xi_1,\xi_2';\xi_2,\xi_2')\equiv \langle\xi_1,\xi_2|\underline{\Gamma}^{(4)}|\xi_2,\xi_2'\rangle$ by
\begin{align}
\Gamma^{(3)}(\xi_1;\xi_2,\xi_2')\equiv&\, \Psi(\xi_1')(-1)^{j_1+j_1'-1}\Gamma^{(4)}(\xi_1,\xi_1';\xi_2,\xi_2')
\notag \\
=&\, \Gamma^{(4)}(\xi_2',\xi_2;\xi_1',\xi_1) (-1)^{j_1+j_1'-1}\Psi(\xi_1')
\notag \\
\equiv &\,  \Gamma^{(3){\rm T}}(\xi_2',\xi_2;\xi_1).
\label{Gamma^(3)-def}
\end{align}
\end{subequations}
Equations (\ref{Gamma^(4)-eq}) and (\ref{Gamma^(3)-def}) are expressible diagrammatically as Fig.\ \ref{fig:2}.

Fourth, $\Gamma^{(3)}$ is connected with the anomalous self-energy by Eq.\ (\ref{n=2-identity}), i.e., 
\begin{align}
\Sigma_{jj}(x,x')=\frac{1}{2}\Psi_{j_1}(x_1)(-1)^{j_1+j-1}\Gamma^{(3)}_{j_1jj}(x_1;x,x') .
\label{n=2-identity-2}
\end{align}
Combined with Eq.\ (\ref{Sigma-Gamma2}), we can conclude that the anomalous self-energy of homogeneous systems vanish in the low energy-momentum
limit due to processes (g) and (h) of Fig.\ \ref{fig:1}, as first shown by Nepomnyashchi\u{i} and Nepomnyashchi\u{i}\cite{Nepomnyashchii75,Nepomnyashchii78}.

\subsection{Equation for ${\Gamma}^{(4{\rm i})}$}

Equations (\ref{DB-GP})--(\ref{n=2-identity-2}) are formally exact, 
which still include the irreducible four-point vertices $\underline{\Gamma}^{(4{\rm i})}$ as unknown functions.
Hence, it is necessary for performing practical microscopic studies to supplement them with equations to determine $\underline{\Gamma}^{(4{\rm i})}$. 
Incidentally, we seek an alternative possibility in the following paper \cite{Kita20-2} of constructing phenomenological parameters in terms of $\underline{\Gamma}^{(4{\rm i})}$
to describe low-energy properties, like the Landau theory of Fermi liquids \cite{Landau56,Landau57,Landau58,SR83}.

To derive the equations for $\underline{\Gamma}^{(4{\rm i})}$, we approximate the functional $\Phi$
in the conserving-gapless form of satisfying Eq.\ (\ref{dPhi}) \cite{Kita09,Kita14}.
To be specific, our  $\Phi$ is given in terms of an unknown effective two-body potential $\tilde{\cal U}(x_1,x_2)\!=\!\tilde{\cal U}(x_2,x_1)$
by \cite{Kita14} 
\begin{align}
\Phi = &\,\frac{1}{2}\tilde{\cal U}(x_1,x_2)\left[\rho(x_1,x_1)\rho(x_2,x_2)+\rho(x_1,x_2)\rho(x_2,x_1)\right.
\notag \\
&\, \left.-\tilde\rho_{11}(x_1,x_2)\tilde\rho_{22}(x_2,x_1)\right] ,
\end{align}
where $\rho$ and $\tilde\rho_{jj}$ are defined by
\begin{subequations}
\begin{align}
\rho(x_1,x_2)\equiv \Psi_1(x_1)\Psi_2(x_2)-\frac{G_{12}(x_1,x_2)+G_{21}(x_2,x_1)}{2},
\end{align}
\begin{align}
\rho_{jj}(x_1,x_2)\equiv \Psi_j(x_1)\Psi_j(x_2)+\frac{G_{jj}(x_1,x_2)+G_{jj}(x_2,x_1)}{2}.
\end{align}
\end{subequations}
The irreducible vertices $\underline{\Gamma}^{(4{\rm i})}$ are obtained from this functional by
Eq.\ (\ref{Gamma^(4i)}).
The basic finite elements are given by
\begin{subequations}
\begin{align}
\Gamma^{(4{\rm i})}_{12;12}(x_1,x_2;x_1',x_2')
=&\, \tilde{U}(x_1,x_1')\left[\delta(x_1,x_2)\delta(x_1',x_2')\right.
\notag \\
&\,\left.+\delta(x_1,x_2')\delta(x_2,x_1')\right],
\label{Gamma_1212}
\\
\Gamma^{(4{\rm i})}_{11;22}(x_1,x_2;x_1',x_2')
=&\, -\tilde{U}(x_1,x_2)\left[\delta(x_1,x_1')\delta(x_2,x_2')\right.
\notag \\
&\, \left.+\delta(x_1,x_2')\delta(x_2,x_1')\right].
\label{Gamma_1122}
\end{align}
\end{subequations}
The other finite elements can be found easily by using the symmetries
$\Gamma^{(4{\rm i})}(\xi_1,\xi_2;\xi_1',\xi_2')\!=\!\Gamma^{(4{\rm i})}(\xi_2,\xi_1;\xi_1',\xi_2')\!=\!\Gamma^{(4{\rm i})}(\xi_1,\xi_2;\xi_2',\xi_1')
\!=\!\Gamma^{(4{\rm i})}(\xi_2',\xi_1';\xi_2,\xi_1)$.
The corresponding $\Gamma^{(3{\rm i})}$ is obtained by Eq.\ (\ref{Gamma^(3i)-def}).
The finite elements are given by
\begin{align}
&\,\Gamma^{(3{\rm i})}_{j;j,3-j}(x_1;x_1',x_2)=\Gamma^{(3{\rm i})}_{3-j;jj}(x_2;x_1,x_1')
\notag \\
=&\, \tilde{U}(x_1,x_1')\left[\delta(x_1,x_2)\Psi_{3-j}(x_1')+\delta(x_1',x_2)\Psi_{3-j}(x_1)\right] .
\end{align}
Thus, Eqs.\ (\ref{Gamma_1212}) and (\ref{Gamma_1122}) yield an identical expression, as they should.

We determine the unknown function $\tilde{U}(x_1,x_1')$ so as to satisfy Eq.\ (\ref{n=2-identity-2}).
Noting that $\Sigma_{22}(x,x')=\Sigma_{11}^*(x,x')$ holds,
we realize that the number of unknown variables, i.e., $\tilde{U}(x,x')$, is equal to the number of constraints 
to be satisfied, i.e., Eq.\ (\ref{n=2-identity-2}). Especially in the weak-coupling cases, we can impose the condition
that $\tilde{U}(x,x')$ approaches the bare interaction potential $U(x,x')$ in the high energy-momentum limit.

\section{Extension to Nonequilibrium Systems\label{sec6}}

The formulation of Sects.\ \ref{sec2}-\ref{sec4} can be extended to 
nonequilibrium systems by (i) performing the inverse Wick rotation $\tau=i t/\hbar$ and (ii) changing the Matsubara contour
$\tau\in[0,\beta]$ into the round-trip Keldysh contour $C$ that extends over $t\in [-\infty,\infty]$ \cite{Keldysh64,Kita10-2}.
We sketch it with (a) modifying the definitions of functions 
and (b) transforming every integral on $C$ into that over $t\in [-\infty,\infty]$ in the second half.

\subsection{Equations on $C$\label{subsec:C}}

We introduce the partition function by
\begin{align}
Z_{JI}
\equiv&\,  \int D[\psi] \, \exp\left\{\frac{i}{\hbar}\left[S_C-\int d\xi\,\psi (\xi) J(\xi)\right.\right.
\notag \\
&\,
\left.\left.-\int d\xi\int d\xi'\psi(\xi)\psi(\xi') I(\xi',\xi)\right]\right\} .
\label{Z_C}
\end{align}
Here $S_C $ is obtained from Eq.\ (\ref{S}) by $S_C\equiv i\hbar S(\tau=it/\hbar)$ with 
an adiabatic factor on $S_{\rm int}$ \cite{Kita10-2}, variable $\xi$ now denotes $\xi\equiv (j,x)$ with $x\equiv ({\bf r},t)$,
and every time integral extends over $C$. 
It satisfies
\begin{subequations}
\label{dlnZ_C}
\begin{align}
i\hbar\frac{\delta \ln Z_{JI}}{\delta J(\xi)}=&\,\langle \psi(\xi)\rangle_{JI}\equiv \Psi_{JI}(\xi),
\label{dlnZ1_C}
\\
i\hbar\frac{\delta \ln Z_{JI}}{\delta J(\xi)\delta J(\xi')}
= & -\!\frac{i}{\hbar}\left[\langle T_C \psi(\xi)\psi(\xi')\rangle_{JI}-\Psi_{JI}(\xi)\Psi_{JI}(\xi')\right]
\notag \\
\equiv& \,G_{JI}(\xi,\xi'),
\label{dlnZ11_C}
\\
i\hbar\frac{\delta \ln Z_{JI}}{\delta I(\xi',\xi)}=&\,\langle T_C \psi(\xi)\psi(\xi')\rangle_{JI}
\notag \\
=&\,  
i\hbar G_{JI}(\xi,\xi')+\Psi_{JI}(\xi)\Psi_{JI}(\xi') 
\notag \\
\equiv  &\,i\hbar\,{\cal G}_{JI}(\xi,\xi'),
\label{dlnZ2_C}
\end{align}
\end{subequations}
where $T_{C}$ arranges operators according to their chronological order on $C$ from right to left. 

Let us perform the Legendre transformation \cite{Weinberg96,dDM64,JL64,Swanson92,NO88}
\begin{align}
\Gamma_{JI}\equiv&\, i\hbar \ln Z_{JI} -\int  d\xi\,\Psi_{JI}(\xi) J(\xi)
\notag \\
&\, -\int  d \xi\int d\xi'\,i\hbar \,{\cal G}_{JI}(\xi,\xi')I(\xi',\xi).
\label{Gamma_C}
\end{align}
Using it, we can follow every step from Eq.\ (\ref{Gamma-deriv}) to Eq.\ (\ref{n=2-identity}),
where the only modification necessary is to
add the factor $i/\hbar$ on the left-hand side of Eq.\ (\ref{Gamma-deriv2}).

We now express $\Gamma$ in terms of another functional $\Phi$,
\begin{align}
\Gamma =&\, \frac{1}{2}\vec\Psi^{\rm T}\hat\sigma_3 \hat{G}_{0}^{-1}\hat\sigma_3\vec\Psi
\notag \\
&\, -\frac{i\hbar}{2}{\rm Tr} \,\left\{\ln\! \left[\bigl(-i\hat\sigma_2\bigr)\bigl(\hat{G}_{0}^{-1}-\hat\Sigma\bigr)\right]+\hat\Sigma\hat{G}\right\}
+\Phi,
\label{LW_C}
\end{align}
similarly as Eq.\ (\ref{LW}) for equilibrium systems.
Accordingly, Eq.\ (\ref{dPhi}) is replaced by
\begin{subequations}
\label{dPhi_C}
\begin{align}
\frac{\partial\Phi}{\partial\Psi(\xi)}=&\, \int d\xi' \,\Sigma(\xi,\xi')(-1)^{j+j'-1}\Psi(\xi'),
\label{dPhi1_C}
\\
\frac{\partial\Phi}{\partial G(\xi',\xi)}=&\,\frac{i\hbar}{2}\Sigma(\xi,\xi') .
\label{dPhi2_C}
\end{align}
\end{subequations}

To derive two-particle Green's functions, we can also follow every step from Eq.\ (\ref{DB-HP}) to Eq.\ 
(\ref{dSigma}).
Equation (\ref{dSigma-Gamma}) is then modified into
\begin{align}
\delta\Sigma(\xi_1,\xi_1')=&\,\frac{i\hbar}{2}\Gamma^{(4{\rm i})}(\xi_1,\xi_1';\xi_2,\xi_2')\delta G(\xi_2,\xi_2') 
\notag \\
&\,+\Gamma^{(3{\rm i}){\rm T}}(\xi_1,\xi_1';\xi_2)\delta \Psi(\xi_2) ,
\label{dSigma-Gamma_C}
\end{align}
where $\Gamma^{(4{\rm i})}$ is now defined by
\begin{align}
\Gamma^{(4{\rm i})}(\xi_1,\xi_1';\xi_2,\xi_2')\equiv  &\,\frac{4}{(i\hbar)^2}\frac{\partial^2\Phi}{\partial G(\xi_1',\xi_1)\partial G(\xi_2,\xi_2')},
\label{Gamma^(4i)_C}
\end{align}
and $\Gamma^{(3{\rm i}){\rm T}}(\xi_1,\xi_1';\xi_2)\equiv \delta\Sigma(\xi_1,\xi_1')/\delta\Psi(\xi_2)$ is expressible in terms of this $\Gamma^{(4{\rm i})}$ as Eq.\ (\ref{Gamma^(3i)-def}).
With the difference between Eqs.\ (\ref{dSigma-Gamma}) and (\ref{dSigma-Gamma_C}) in mind,
we can proceed in exactly the same way as from Eq.\ (\ref{VecMat}) to Eq.\ (\ref{dPsidG-sol}).
Indeed, the only modification necessary is to replace every prefactor ${1}/{2}$ by $-{i\hbar}/{2}$.
Thus, Eq.\ (\ref{dPsidG-sol}) is replaced by
\begin{subequations}
\label{dPsidG-sol_C}
\begin{align}
\delta\vec{\Psi}
=&\,\hat{G}\biggl(\underline{\Psi}^{(3)}
+\frac{i\hbar}{2}\,\underline{\Gamma}^{(3)}\underline{GG} \biggr)\delta \vec{I}^{\,({\rm s})}  ,
\label{dPsi-sol_C}
\\
\delta\vec{G}=&\,\left[\underline{GG}\left(\underline{1}+\frac{i\hbar}{2}\,\underline{\Gamma}^{(4)}\underline{GG}\right) 
\right.
\notag \\
&\,\left. +\,\underline{GG}\,\underline{\Gamma}^{(3){\rm T}}\,\hat{G}\biggl(\underline{\Psi}^{(3)}
+\frac{i\hbar}{2}\,\underline{\Gamma}^{(3)}\underline{GG} \biggr)\right]\delta \vec{I}^{\,({\rm s})} .
\label{dG-sol_C}
\end{align}
\end{subequations}

Let us introduce the $n$-point Green's functions by
\begin{align}
G^{(n)}(\xi_1,\cdots,\xi_n)
=&\,(i\hbar)^{n-\left[\frac{n}{2}\right]}\frac{\delta^n \ln Z_J}{\delta J(\xi_1)\cdots J(\xi_n)}\Biggr|_{J=0}.
\label{G^(n)-def_C}
\end{align}
The first two functions are expressible in terms of those given 
in Eq.\ (\ref{dlnZ_C}) by $G^{(1)}(\xi)=\Psi(\xi)$ and $G^{(2)}(\xi_1,\xi_2)=G(\xi_1,\xi_2)$. 
Accordingly, Eq.\ (\ref{G^(3,4)-2}) is modified into
\begin{subequations}
\label{G^(3,4)-2_C}
\begin{align}
G^{(3)}_{123}=&\,\frac{\delta\Psi_1}{\delta I_{32}}\Biggr|_{I=0}
-\Psi_2G_{13}-\Psi_3G_{12} ,
\label{G^(3,4)-21_C}
\\
G^{(4)}_{1234}
=&\,\frac{\delta G_{12}}{\delta I_{43}}\Biggr|_{I=0}+\frac{i}{\hbar}\left(\Psi_3G^{(3)}_{124}
+\Psi_4G^{(3)}_{123}\right)
\notag \\
&\, -G_{13}G_{24}-G_{14}G_{23} .
\label{G^(3,4)-22_C}
\end{align}
\end{subequations}
Substituting Eq.\ (\ref{dPsidG-sol_C}) into Eq.\ (\ref{G^(3,4)-2_C}), we obtain
\begin{subequations}
\label{G^(3,4)_C}
\begin{align}
&\,G^{(3)}(\xi_1,\xi_2,\xi_3)
\notag \\
=&\,i\hbar \,G(\xi_1,\xi_1')G(\xi_2,\xi_2')G(\xi_3,\xi_3')\Gamma^{(3)}(\xi_1';\xi_2',\xi_3') ,
\label{G^(3)_C}
\\
&\,G^{(4)}(\xi_1,\xi_2,\xi_3,\xi_4)
\notag \\
=&\,i\hbar\,G(\xi_1,\xi_1')G(\xi_2,\xi_2')G(\xi_3,\xi_3')G(\xi_4,\xi_4')
\notag \\
&\,\times 
\left[\Gamma^{(4)}(\xi_1',\xi_2';\xi_3',\xi_4')
+\Gamma^{(3){\rm T}}(\xi_1',\xi_2';\xi_5) G(\xi_5,\xi_5')\right.
\notag \\
&\,\left.\times \Gamma^{(3)}(\xi_5';\xi_3',\xi_4')
\right]
 .
\label{G^(4)_C}
\end{align}
\end{subequations}

Expressions of the self-energies can also be obtained 
by following the procedure of Sect.\ \ref{sec4}.
We thereby obtain
\begin{align}
&\,\Sigma(\xi_1,\xi_1')
\notag \\
=&\, \delta_{j_13-j_1'}\delta(x_1,x_1') U(x_1,x_2)
\bigl[\Psi_2(x_2)\Psi_1(x_2)
\notag \\
&\, +i\hbar \,G_{21}(x_2,x_2)\bigr]
+U(x_1,x_1')\bigl[\Psi_{3-j_1}(x_1)\Psi_{3-j_1'}(x_1')
\notag \\
&\, +i\hbar \,G_{3-j_13-j_1'}(x_1,x_1')\bigr]
\notag \\
&\, +\frac{1}{2} U(x_1,x_2)\bigl\{(i\hbar)^2 G_{3-j_1j_3}(x_1,x_3) G_{2j_4}(x_2,x_4)
\notag \\
&\, \times G_{1j_4'}(x_2,x_4')\bigl[\Gamma^{(4)}(\xi_3,\xi_1';\xi_4,\xi_4')+\Gamma^{(3){\rm T}}(\xi_3,\xi_1';\xi_5)
\notag \\
&\,\times G(\xi_5,\xi_5')\Gamma^{(3)}(\xi_5';\xi_4,\xi_4')\bigr]
\notag \\
&\,+i\hbar G_{3-j_1j_3}(x_1,x_3)\bigl[
G_{2j_4}(x_2,x_4)\Psi_{1}(x_2)
\notag \\
&\,\hspace{23mm}+\Psi_{2}(x_2)G_{1j_4}(x_2,x_4)\bigr]
 \Gamma^{(3){\rm T}}(\xi_3,\xi_1';\xi_4)
\notag \\
&\, +i\hbar \Psi_{3-j_1}(x_1)G_{2j_4}(x_2,x_4)G_{1j_4'}(x_2,x_4')\Gamma^{(3)}(\xi_1';\xi_4,\xi_4') 
\notag \\
&\, +(\xi_1\leftrightarrow\xi_1')\bigr\}.
\label{Sigma-Gamma_C}
\end{align}

\subsection{Equations on $t\in[-\infty,\infty]$}

Every integral over $C$ can be transformed into that of $t\in[-\infty,\infty]$ through
the procedure \cite{Keldysh64,Kita10-2}
\begin{align}
\int_{-\infty}^\infty dt^C =\int_{-\infty}^\infty dt^1+\int_{\infty}^{-\infty} dt^2
=\int_{-\infty}^\infty dt^1-\int_{-\infty}^{\infty} dt^2,
\end{align}
where $t^1$ ($t^2$) denotes a time on the outward (return) path. We also introduce
the argument $x^i =({\bf r},t^i)$ $(i=1,2)$, and the corresponding Green's functions and self-energies by
\begin{subequations}
\begin{align}
G_{j_1j_2}^{i_1i_2}(x_1,x_2)\equiv &\, G_{j_1j_2}(x_1^{i_1},x_2^{i_2}),
\\
\Sigma_{j_1j_2}^{i_1i_2}(x_1,x_2)\equiv&\, (-1)^{i_1+i_2}\Sigma_{j_1j_2}(x_1^{i_1},x_2^{i_2}) .
\label{Sigma^ii'}
\end{align}
\end{subequations}
Let us define the matrices in the Nambu-Keldysh space (i.e., the $j$-$i$ space) by
\begin{subequations}
\begin{align}
\check{G}(x,x')\equiv \begin{bmatrix} \hat{G}^{11}(x,x') & \hat{G}^{12}(x,x') \\
\hat{G}^{21}(x,x') & \hat{G}^{22}(x,x')
\end{bmatrix},
\end{align}
\begin{align}
\check\Sigma(x,x')\equiv \begin{bmatrix} \hat\Sigma^{11}(x,x') & \hat\Sigma^{12}(x,x') \\
\hat\Sigma^{21}(x,x') & \hat\Sigma^{22}(x,x')
\end{bmatrix},
\end{align}
\begin{align}
\check{G}_0^{-1}(x,x')\equiv \begin{bmatrix} \hat{G}_0^{-1}(x,x') & 0 \\
0 & -\hat{G}_0^{-1}(x,x')
\end{bmatrix},
\label{checkG_0^-1}
\end{align}
\begin{align}
\check{1}\equiv \begin{bmatrix} \hat{1} & \hat{0} \\
\hat{0} & \hat{1}
\end{bmatrix},\hspace{10mm}
\check{\sigma}_3\equiv \begin{bmatrix} \hat{\sigma}_3 & \hat{0} \\
\hat{0} & \hat{\sigma}_3
\end{bmatrix}, 
\label{check1}
\end{align}
\end{subequations}
where each quantity with a hat on it is a $2\times 2$ matrix in the Nambu space, e.g., $\hat{G}_0^{-1}(x,x')$ is
given by Eq.\ (\ref{hatG_0^-1}) with $\tau=it/\hbar$.
The factor $(-1)^{i_1+i_2}$ in Eq.\ (\ref{Sigma^ii'}) and the $-$ sign in the 22 element of Eq.\
(\ref{checkG_0^-1}) have been added to our previous definitions \cite{Kita10-2}.
Because of these modifications, $\check{G}$ now obeys 
\begin{align}
(\check{G}_0^{-1}-\check{\Sigma})\check{G}=\check{1} .
\end{align}
Similarly, the generalized Hugenholtz-Pines relation is given by
\begin{align}
\check\sigma_3(\check{G}_0^{-1}-\check{\Sigma})\check\sigma_3\vec{\Psi}=\vec{0} ,
\end{align}
where $\vec{\Psi}$ and $\vec{0}$ are vectors with four elements with 
\begin{align}
\vec\Psi(x)\equiv \begin{bmatrix}\vec\Psi(x^1) \\ \vec\Psi(x^2) \end{bmatrix} .
\end{align}

With these quantities, derivation of the two-particle Green's functions on $t\in [-\infty,\infty]$
proceeds in exactly the same way as that of Sect.\ \ref{subsec:C}.
Indeed, we only need to redefine $\xi$ by $\xi\equiv (i,j,x)$, replace every $\hat{A}$ by $\check{A}$,
and use
\begin{align}
\delta\check{I}(x,x')\equiv \begin{bmatrix} \delta\hat{I}(x,x') & \hat{0} \\
\hat{0} & -\delta\hat{I}(x,x')
\end{bmatrix},
\label{checkI}
\end{align}
in place of $\delta\hat{I}(x,x')$. Equation (\ref{dPsidG-sol_C}) is thereby replaced by
\begin{subequations}
\label{dPsidG-sol_K}
\begin{align}
\delta\vec{\Psi}
=&\,\check{G}\biggl(\underline{\Psi}^{(3)}
+\frac{i\hbar}{2}\,\underline{\Gamma}^{(3)}\underline{GG} \biggr)\delta \vec{I}^{\,({\rm s})}  ,
\label{dPsi-sol_K}
\\
\delta\vec{G}=&\,\left[\underline{GG}\left(\underline{1}+\frac{i\hbar}{2}\,\underline{\Gamma}^{(4)}\underline{GG}\right) 
\right.
\notag \\
&\,\left. +\,\underline{GG}\,\underline{\Gamma}^{(3){\rm T}}\,\check{G}\biggl(\underline{\Psi}^{(3)}
+\frac{i\hbar}{2}\,\underline{\Gamma}^{(3)}\underline{GG} \biggr)\right]\delta \vec{I}^{\,({\rm s})} .
\label{dG-sol_K}
\end{align}
\end{subequations}

\section{Concluding Remarks\label{sec7}}

We have derived a system of equations for the condensate wave function, Green's functions, and three- and four-point Green's functions,
which are summarized in Sect.\ \ref{sec5}. The four-point (i.e., two-particle) Green's functions are confirmed to share poles with 
Green's functions, as seen in Eq.\ (\ref{G^(4)}). 
One of the other key results here is Eq.\ (\ref{dPsidG-sol}), which will be used in the following paper \cite{Kita20-2} to derive the 
Ward-Takahashi identities. They will enable us to describe low-temperature properties of correlated Bose-Einstein condensates
in terms of low-energy Green's functions and vertices.

\section*{Acknowledgment}
This work was supported by JSPS KAKENHI Grant Number JP20K03848.

\end{document}